\documentclass[a4paper,noarxiv,twocolumn,11pt, ]{quantumarticle}
\pdfoutput=1

\usepackage{graphicx}
\usepackage{amsfonts}
\usepackage{amsmath}

\usepackage[numbers,sort&compress]{natbib}

\usepackage[usenames,dvipsnames]{color}
\usepackage{bm}
\usepackage{amssymb}
\usepackage{braket}

\usepackage{multirow}
\usepackage{epstopdf}
\usepackage[normalem]{ulem}

\usepackage{hyperref}
\hypersetup{
colorlinks=true,
linkcolor=blue,
citecolor=blue,
filecolor=green,
urlcolor=blue,
}

\usepackage[caption=false]{subfig}

\begin{document}

\title{Bringing multilevel quantum master equations into Lindblad form for complete positivity tests: Two approaches}
%
\author{Timur V. Tscherbul}
\affiliation{Department of Physics, University of Nevada, Reno, Nevada, 89557, USA}

\begin{abstract}

While quantum master equations (QMEs) are the primary workhorse in quantum information science, quantum optics, spectroscopy, and quantum thermodynamics, bringing an arbitrary $N$-level QME into Lindbladian form and verifying complete positivity of the associated quantum dynamical map remain open challenges for $N\ge 3$. We explore and implement  two independent methods to accomplish these tasks, which enable one to directly compute the Kossakowski matrix of an arbitrary Markovian QME from its Liouvillian.
In the first method, due to Hall, Cresser, Li, and  Andersson, 
the Kossakowski matrix elements are obtained by  evaluating the action of the Liouvillian on the orthonormal SU($N$) basis matrices and then computing a sum of matrix-product traces.  The second method, developed in this work, is based on the real $N$-level coherence vector and relies on the Moore-Penrose pseudo-inverse of a rectangular matrix composed of the structure constants of SU$(N)$.  We show that both methods give identical results, and  apply them  
to establish the complete positivity of the partial secular Bloch-Redfield QME for the $\Lambda$ and V-systems driven by incoherent light.  We find that the  eigenvalues of the Kossakowski matrix of these seemingly different three-level systems are identical,  implying close similarities of their dissipative dynamics.
  By facilitating the expression of multilevel Markovian QMEs in Lindblad form, our results enable testing the QMEs for  complete positivity without solving them, as well as restoring complete positivity  by keeping only non-negative eigenvalues of the Kossakowski matrix.   

 \end{abstract}
\maketitle

\section{Introduction}

Quantum master equations (QMEs) describe the time evolution of a quantum subsystem of interest coupled to a (typically) much larger environment \cite{AlickiBook,Breuer:06,Nielsen:10}. As such, they play a vital role in many areas of physics and chemistry, where the accurate description of environmental effects  on quantum dynamics is essential, such as quantum information science \cite{Campaioli:24, McCauley:20,Davidovic:20,Mozgunov:20,Manzano:20,Groszkowski:23,Albert:14,Albert:16,Abbruzzo:23,Abbruzzo:24prx,Pradilla:24}, quantum optics \cite{Breuer:06,Scully:97}, quantum thermodynamics \cite{QuantumThermodynamicsBook,Alicki:79,Kosloff:19,Potts:24,Liu:21,Scali:21,Myers:22,Soret:22,Onishchenko:24},  spectroscopy \cite{YuenZhou:11,Mancal:12,YuenZhouBook,Fetherolf:17, Sayer:24}, and chromophoric energy transport  \cite{Rebentrost:09,Cao:09,Gelin:11,Leon-Montiel:14,Tscherbul:18}. When used within their respective domains of validity, QMEs offer a robust and physically meaningful  description of environment-induced relaxation and decoherence in $N$-level quantum systems \cite{AlickiBook,Nielsen:10,Breuer:06}.

A crucial property of any QME is complete positivity (CP) of the quantum dynamical map associated with it \cite{AlickiBook,Nielsen:10,Breuer:06}. The CP condition requires not only that the reduced dynamics of a subsystem of interest be positive, but also that it remains so when the subsystem is entangled with arbitrary ancillae \cite{AlickiBook,Nielsen:10,Breuer:06}.
This condition is a defining property of a quantum channel \cite{Nielsen:10} and its violation can lead to unphysical effects, such as entanglement generation by purely local interactions \cite{Benatti:02,Benatti:22} and violations of the second law of thermodynamics \cite{Argentieri:14,Soret:22}. Given recent advances in quantum thermodynamics \cite{Myers:22,Onishchenko:24} and quantum information processing \cite{Wang:20,Chi:22} with  multilevel quantum systems (qudits), the question of whether the dynamical map generated by a given QME satisfies the CP condition acquires critical importance.

In principle,
this question can be resolved by mapping the QME's  Liouvillian matrix (expressed in terms of  relaxation and decoherence rates) to the so-called Kossakowski matrix $\mathbf{A}$, which parametrizes the canonical Gorini-Kossakowski-Lindblad-Sudarshan (GKLS) generator of the quantum dynamical semigroup associated with the QME \cite{AlickiBook}. The dynamical map is CP if $\mathbf{A}$  has only nonnegative eigenvalues  \cite{AlickiBook}. 
The Kossakowski matrix determines the key properties of the quantum dynamical map, and plays an essential role in, e.g., quantifying the extent of non-Markovianity of quantum evolutions  \cite{Rivas:10, Abbruzzo:23}.

While  the analytical mapping between the Liouvillian and   Kossakowski matrices is well-established for the two-level system  \cite{AlickiBook}, this is not the case for multilevel quantum systems ($N\ge3$).
Although analytic expressions for the relaxation and decoherence rates in terms of the elements of the Kossakowski matrix were obtained for $N=3$ \cite{PottingerThesis}, these expressions are quite cumbersome \cite[p.~69]{AlickiBook} and have not been inverted to the best of our knowledge.
As a result,  Kossakowski matrices for $N\ge 3$ are presently calculated via indirect methods, which are applicable  
only to specific subclasses of QMEs, such as  the secular QME \cite{AlickiBook} and certain nosecular QMEs of the Bloch-Redfield type, both Markovian and non-Markovian \cite{Farina:19,Trushechkin:21,McCauley:20,Davidovic:20,Vacchini:16}.  These QMEs are typically expressed in operator form,
with the Kossakowski matrices expressed via Fourier transforms of environmental correlation functions.
  However, a limitation of these approaches is their lack of generality, i.e., it is not always clear whether  a given QME can be expressed in one of the prescribed operator forms.   
 Therefore, obtaining the Kossakowski matrix  from the Liouvillian of an arbitrary multilevel QME remains an open challenge  \cite{AlickiBook}, preventing robust CP testing of the underlying quantum dynamical maps.

In 2007, Andersson, Cresser, and Hall  showed  that  any time-local QME can be decomposed into Kraus operators, and thereby tested for CP by constructing the Choi matrix associated with its Kraus-type representation  \cite{Andersson:07}. Specifically, a quantum dynamical map is CP if and only if the corresponding Choi matrix is positive \cite{Choi:75,Andersson:07}.  This method is applicable to a wide class of QMEs, including  non-Markovian QMEs written in a time-local form (which can be obtained from the generalized Nakajima-Zwanzig-type QME via the time-convolutionless projection operator method \cite{Breuer:01,Shibata:77,Chaturvedi:79,Caves:99}).
However, the construction of the Choi matrix and the associated  Kraus decomposition requires explicitly solving the QME \cite{Andersson:07}, which can be a computationally intensive task for multilevel QMEs. 
As such, its applications  have been limited to two-level systems  \cite{Andersson:07}.

More recently,  Hall, Cresser, Li, and Andersson \cite{Hall:14} pointed out that the Kossakowski matrix can be obtained by evaluating the action of the Liouvillian on the basis matrices of  SU($N$), and then computing the trace of a matrix product,  as given by Eq.~(A7) of Ref. \cite{Hall:14}.
However, the computational significance of this result has remained unexplored. In particular,  to the author's knowledge, it has not been applied to calculate the Kossakowski matrices of multilevel ($N\ge 3$) systems. 

Here, we show that the approach  of Ref. \cite{Hall:14} can be used as the basis for an efficient method to calculate the Kossakowski matrices of $N$-level Markovian QMEs  from their  Liouvillians. Focusing on the quantum optical QME of the partial secular Bloch-Redfield (PSBR) type \cite{Kozlov:06,Tscherbul:14,Tscherbul:15b}, which describes  multilevel atoms and molecules interacting with incoherent radiation fields, we  use the method to calculate the Kossakowski matrix of the PSBR equation for three-level V and $\Lambda$  systems and show that the corresponding dynamical map is CP.  In addition to the method of Ref.~\cite{Hall:14}, which is based on the density matrix formalism,  we develop an alternative approach, which relies on the real coherence vector, and show that both approaches give identical results.

The quantum dynamics of multilevel atoms and molecules interacting with incoherent radiation fields (such as sunlight, or, more generally, thermal blackbody radiation) is of paramount importance in a number of research fields, including precision measurement \cite{Beloy:06,Dodin:24},  photosynthetic light-harvesting \cite{Kassal:13,Tscherbul:14,Dodin:16,Dodin:18,Koyu:18,Koyu:21,Koyu:22,Dodin:24,Brumer:18}, and quantum thermodynamics \cite{Scully:11,Dorfman:13}. 
Interestingly, these dynamics  can generate the so-called noise-induced Fano coherences between the nearly degenerate ground and/or excited states of a multilevel quantum system \cite{Agarwal:01,Kozlov:06,Tscherbul:14,Tscherbul:15b,Dodin:16,Dodin:18,Koyu:18,Koyu:21,Koyu:22,Scully:11,Dorfman:13,Dodin:24,Brumer:18}, even when the initial state of the system is coherence-free.
The emergence of Fano coherences under incoherent driving is due to the non-secular population-to-coherence coupling terms properly described by the PSBR equation  \cite{Tscherbul:15b,Dodin:16,Dodin:18,Koyu:18,Koyu:21,Koyu:22} but not the simpler secular rate equations. The physical origin of these non-secular terms can be traced back to quantum interference between the different incoherent excitation and relaxation pathways \cite{Dodin:16,Koyu:21}.

Far from a mere mathematical curiosity, Fano coherences have a number of experimentally verifiable \cite{Dodin:18,Koyu:22} physical implications. They can  strongly affect the thermalization dynamics of a multilevel quantum system interacting with a bath  \cite{Tscherbul:14,Koyu:18,Koyu:22}, leading to anomalously slow approach to thermal equilibrium which occurs over multiple, often vastly different, timescales \cite{Tscherbul:14,Dodin:16,Dodin:18,Koyu:18,Koyu:21,Koyu:22,Merkli:15,Ivander:23,Gerry:24}. The resulting coherent steady states can break detailed balance \cite{Kozlov:06,Scully:11,Dodin:18,Koyu:21} thereby enhancing the  efficiency of photovoltaic devices \cite{Scully:11}. In addition, closely related pre-thermal states can have promising applications in quantum thermometry  \cite{AntoSztrikacs:24}.

Although the PSBR equation is the main tool, with which non-secular effects and Fano coherences are modeled,  
 it has been an open question whether the time evolution generated by this equation satisfies the CP property  \cite{Tscherbul:14,Trushechkin:21}, calling into question the physical significance of non-secular effects and Fano coherences.
 Here, we settle this question by using the method of Ref.~\cite{Hall:14} to explicitly calculate the Kossakowski matrix for the archetypal three-level $\Lambda$ and V-systems. We show that the eigenvalues of the Kossakowski matrix are non-negative,  establishing the CP property of the PSBR dynamical map acting on these systems. Our calculations further show that the eigenvalues of the Kossakowski matrices of the $\Lambda$ and V-systems are identical, implying close similarities in the dissipative dynamics of these seemingly different systems.

\section{Theoretical approaches: Extracting the Kossakowski matrix from the Liouvillian}

We begin by specifying the general Markovian QME for the reduced  density matrix $\bm{\rho}$ of an $N$-level quantum system described by the Hamiltonian $\mathbf{H}_S$ interacting with an environment \cite{AlickiBook,Breuer:06}
\begin{equation}\label{QME}
{d}{\bm{\rho}}/dt =\hat{\mathcal{L}}[\bm{\rho}] = \hat{\mathcal{L}}_H[\bm{\rho}] + \hat{\mathcal{L}}_D[\bm{\rho}],
\end{equation}
 where the Liouvillian superoperator $\hat{\mathcal{L}}$, assumed here to be time-independent, can be separated into the Hamiltonian ($\hat{\mathcal{L}}_H[\bm{\rho}]=-i[\mathbf{H}_S,\bm{\rho}]$) and dissipative parts. The latter can be expressed in the GKLS form  \cite{AlickiBook} 
\begin{equation}\label{L_D}
\hat{\mathcal{L}}_D[\bm{\rho}] = \frac{1}{2}\sum_{i,k=1}^{M} a_{ik} \left( 2 \mathbf{F}_i\bm{\rho}\mathbf{F}_k - \bm{\rho}\mathbf{F}_k\mathbf{F}_i - \mathbf{F}_k\mathbf{F}_i\bm{\rho}  \right),
\end{equation}
where $a_{ik}$ are the elements of the Kossakowski matrix $\mathbf{A}$, and the basis set $\{\mathbf{F}_i\}$ is composed of $M=N^2-1$ Hermitian traceless $N\times N$ matrices, which satisfy $\text{Tr}(\mathbf{F}_i\mathbf{F}_k^\dag)=\delta_{ik}$ \cite{AlickiBook}. For $N=3$, the basis matrices $\mathbf{F}_i$ correspond to the generators of the SU$(3)$ Lie group, also known as error generators \cite{Lidar:98}. We follow the standard practice of  choosing the $\mathbf{F}_i$ to be the Gell-Mann matrices \cite{AlickiBook,Macfarlane:68}.  Note that the summation in Eq.~\eqref{L_D} does not include the basis matrix $\mathbf{F}_0$, which is proportional to the unit matrix, and thus commutes with $\hat{\rho}$ and the $\mathbf{F}_k$ ($k=1,M$), so the terms involving  $\mathbf{F}_0$ evaluate to zero.
Our goal is to obtain  the matrix elements of  $\mathbf{A}$ in terms of those of  the Liouvillian $\hat{\mathcal{L}}_D$.

Hall and co-workers derived   the following expression for the Kossakowski matrix  in terms of the Liouvillian   \cite{Hall:14},
which follows from the proof of Lemma 2.2 of Ref.~\cite{Gorini:76}
\begin{equation}\label{aij_DM}
a_{ij} = \sum_{m=0}^{M} \text{Tr}(\mathbf{F}_i \mathbf{F}_m \hat{\mathcal{L}}_D[\mathbf{F}_m]\mathbf{F}_j).
\end{equation}
The calculation of the Kossakowski matrix elements thus requires  evaluating the action of the Liouvillian superoperator on the SU($N$) basis matrices (the term $\hat{\mathcal{L}}_D[\mathbf{F}_m]$)  followed by the computation of  the matrix product trace.

Here, we use Eq.~\eqref{aij_DM} as a basis for calculating the Kossakowski matrices for archetypal three-level V and $\Lambda$ systems driven by incoherent radiation (see Sec. III below).  In principle, however, Eq.~\eqref{aij_DM} is general and can be extended to $N$-level systems with $N\ge 4$ \cite{Dodin:24} and time-dependent Liouvillians.

As an  alternative to Eq.~\eqref{aij_DM}, we propose an independent approach to validate our results. The approach is based on the real coherence vector $\mathbf{v}$ \cite{AlickiBook}, whose components are defined by the SU($N$) decomposition of $\bm{\rho}$ \cite{AlickiBook,Lendi:87,Pottinger:85}
\begin{equation}\label{SUNdecomp}
\bm{\rho}(t) = \frac{1}{N}\mathbf{F}_0 + \sum_{i=1}^M v_i(t) \mathbf{F}_i
\end{equation}
with $v_i(t)=\text{Tr}(\bm{\rho}(t)\mathbf{F}_i)$ ($\mathbf{v}$ is identical to the standard Bloch vector for $N=2$ \cite{AlickiBook}). The real components $v_i$ satisfy an inhomogeneous QME, which is equivalent to the original QME \cite{AlickiBook,Lendi:87,Pottinger:85}
\begin{equation}\label{coherence_vec}
\dot{\mathbf{v}}(t) = \left[\mathbf{Q}+\mathbf{R}\right]\mathbf{v}(t) + \mathbf{k}, 
\end{equation}
where $\mathbf{k}$ is the driving vector and the matrix operators $\mathbf{Q}$ and $\mathbf{R}$ describe the Hamiltonian and dissipative evolution  analogously to $\hat{\mathcal{L}}_H$ and $\hat{\mathcal{L}}_D$ in the density matrix formulation.

We can express the elements of $\mathbf{R}$ and $\mathbf{k}$ in terms of those of $\mathbf{A}$ \cite{AlickiBook,Lendi:87,Pottinger:85}
\begin{align}\label{lineq1_main}
r_{sm}&=\sum_{i,k=1}^M \mathcal{T}_{sm,ik} a_{ik} \quad (s,m=1,\ldots,M) ,\\ \label{lineq2_main}
k_{s}&=\frac{i}{N} \sum_{i,k=1}^M a_{ik} f_{kls} \quad (s=1,\ldots,M),
\end{align}
where
\begin{align}\label{ttensor_main}\notag
\mathcal{T}_{sm,ik} =-\frac{1}{4} &\sum_{l=1}^M [(f_{mnl}+id_{mnl})f_{kls} \\
 &\,\,\, + (f_{klm}-id_{klm}) f_{ils}) ],
\end{align}
is an $M^2\times M^2$ transformation tensor  $\mathcal{T}$, which depends on the symmetric ($f_{mnl}$) and antisymmetric ($d_{mnl}$) structure constants of SU$(N)$ \cite{AlickiBook}. To the author's knowledge, this tensor  has not been explored beyond the two-level system (some of its properties are derived in Appendix~A). 

We note that Eqs.~\eqref{lineq1_main}-\eqref{lineq2_main}  can be viewed as a system of linear equations for the coefficients $a_{ik}$ with the right-hand side given by the $r_{sm}$ and $k_s$. 
This system  can be solved for the $a_{ik}$ as described in the Appendix, to give  
 the vectorized  Kossakowski matrix  $\bm{a} =\text{vec}(\mathbf{A})$ 
\begin{equation}\label{avector_text_main}
\bm{a}=\bm{{T}}^{+} \bm{r}, 
\end{equation}
where $\bm{T}$ is a $(M^2+M) \times M^2$ complex rectangular tensor composed of the structure constants of SU($N$) and 
$\bm{{T}}^{+} = (\bm{T}^\dag \bm{T})^{-1} \bm{T}^\dag$
is the Moore-Penrose pseudo-inverse of $\bm{T}$ \cite{Moore:20,Penrose:55,Barata:12}.

\begin{figure}
\captionsetup[subfigure]{margin={0.05cm,0.0cm},font=normalsize}
    \centering
\subfloat[] {
  \includegraphics[width=0.35\columnwidth, trim = 0 0 0 0]{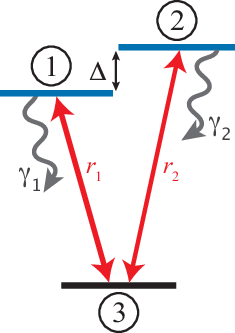}
}
\hspace{0.4cm}
\subfloat[]{
  \includegraphics[width=0.35\columnwidth,trim =  0 2 0 0]{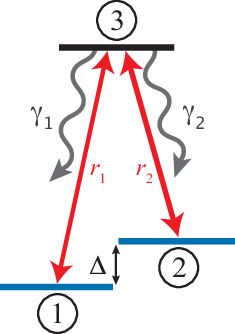}
}	
\caption{(a) Schematics of three-level V (a) and $\Lambda$ (b) systems driven by incoherent radiation. The incoherent pumping transitions are denoted by straight lines  and spontaneous emission transitions by wavy lines. }
	\label{fig:Vsys}
\end{figure}

\section{Results: Complete positivity of Partial Secular Bloch-Redfield evolution of three-level V and $\Lambda$ systems}

We have verified by direct numerical computations that Eqs.~\eqref{aij_DM} and \eqref{avector_text_main} give identical results for the Kossakowski matrix of the three-level V-system. However, the density matrix formulation based on Eq.~\eqref{aij_DM} 
is more computationally efficient because it does not rely on matrix inversion.
Here, we use this formulation to establish the CP property of the partial secular Bloch-Redfield equations, which describe the time evolution of Fano coherences in three-level V and $\Lambda$ systems driven by incoherent radiation.

\subsection{V-system}

The quantum dynamics of a three-level V-system driven by incoherent blackbody radiation [see Fig.~1(a)] is described by the PSBR equations in the energy basis \cite{Tscherbul:15b,Dodin:16,KoyuThesis,Donati:24}
\small
\begin{align}\label{QME}\notag
\dot{\rho}_{i i} &=-(r_i + \gamma_i)\rho_{ii} +r_i \rho_{33}-p(\sqrt{r_1r_2}+\sqrt{\gamma_1\gamma_2})\rho_{12}^R,  \\ 
\dot{\rho}_{1 2}&= -i\rho_{12}\Delta -\frac{1}{2}(r_1+r_2+\gamma_1+\gamma_2)\rho_{12}  \\ \notag
                &+\frac{p}{2}\sqrt{r_1 r_2}(2\rho_{33}-\rho_{1
                  1}-\rho_{2
                  2})-\frac{p}{2}\sqrt{\gamma_1\gamma_2}(\rho_{1
                  1}+\rho_{2 2}), \\ \notag
\dot{\rho}_{13} &= -\frac{1}{2}(2r_1+r_2+\gamma_1)\rho_{13} - \frac{p}{2}(\sqrt{r_1r_2}+\sqrt{\gamma_1\gamma_2})\rho_{23},\\ \notag
\dot{\rho}_{23} &= -\frac{1}{2}(r_1+2r_2+\gamma_2)\rho_{23} - \frac{p}{2}(\sqrt{r_1r_2}+\sqrt{\gamma_1\gamma_2})\rho_{13},
\end{align}
\normalsize
where $i=1,2$ denote the excited states of the V-system [see Fig. 1(a)], $3$ stands for the ground state, $r_i$ and $\gamma_i$ are incoherent absorption and spontaneous emission rates,  $\Delta=\omega_{1 2}$ is the excited-state energy splitting, $p=\bm{\mu}_{3 1}\cdot\bm{\mu}_{3 2}/(|\mu_{3 1}| |\mu_{3 2}|)$ quantifies the alignment of transition dipole moments $\bm{\mu}_{3 i}$, and $\rho_{12}^R$ and $\rho_{12}^I$  denote the real and imaginary parts of $\rho_{12}$ \cite{Kozlov:06,Tscherbul:14,Dodin:16}. Equation~\eqref{QME} is derived in the standard Born-Markov approximation, which is well-justified for multilevel quantum systems interacting with incoherent radiation fields \cite{Breuer:06,Tscherbul:15b}. We retain the nonsecular population-to-coherence coupling terms proportional to $p$ in Eq.~\eqref{QME}, which are responsible for the generation of noise-induced Fano coherences \cite{Kozlov:06,Tscherbul:14}.

To  compute the traces of  matrix products on the right-hand side of Eq.~\eqref{aij_DM},  we need to evaluate the $N\times N$ matrices $\mathcal{L}_D[\mathbf{F}_m]$ for the $N$-level system of interest. To this end, we express the $3\times 3$ density matrix of the V-system as  a complex 9-dimensional vector
\begin{equation}\label{DMvec}
 \vec{\rho}=(\rho_{11}, \rho_{22}, \rho_{33}, \rho_{12}, \rho_{21},\rho_{13},\rho_{31},\rho_{23},\rho_{32})^T, 
 \end{equation}
The  Liouvillian then becomes a $9\times 9$ supermatrix. The time evolution of state populations $\rho_{ii}$ and two-photon coherences $\rho_{12}=\rho_{21}^*$ is decoupled from that of one-photon coherences $\rho_{13}$ and $\rho_{23}$ in the  partial secular approximation, which is well-justified for the V- and $\Lambda$ systems considered here \cite{Tscherbul:15b}. Owing to the partial secular decoupling, the Liouvillian matrix becomes block-diagonal, $\bm{\mathcal{L}}_D = \bm{\mathcal{L}}_D^{(5)}\oplus \bm{\mathcal{L}}_D^{(4)}$ with the $5\times 5$ subblock acting on the top 5 elements of $\vec{\rho}$ ($\rho_{11}$, $\rho_{22}$, $\rho_{33}$, $\rho_{12}$, and  $\rho_{21}$) given by
\begin{widetext}
\begin{equation}\label{LD5}
\bm{\mathcal{L}}_D^{(5)} = 
 \begin{pmatrix}
 -(r_1+\gamma_1) & 0 & r_1 & \xi_{12}(p) & \xi_{12}(p)  \\
 0 & -(r_2+\gamma_2) & r_2 & \xi_{12}(p) & \xi_{12}(p)  \\
r_1+\gamma_1 & r_2+\gamma_2 & -(r_1+r_2) & -2\xi_{12}(p) & -2\xi_{12}(p)  \\
\xi_{12}(p)  & \xi_{12}(p)  & p\sqrt{r_1 r_2} & -\frac{1}{2}(r_1+r_2+\gamma_1+\gamma_2) & 0  \\
\xi_{12}(p)  & \xi_{12}(p)  & p\sqrt{r_1 r_2} & 0 & -\frac{1}{2}(r_1+r_2+\gamma_1+\gamma_2) \\
 \end{pmatrix}
 \end{equation}
 \end{widetext}
where $\xi_{12}(p)=-\frac{p}{2}(\sqrt{r_1 r_2}+\sqrt{\gamma_1\gamma_2})$. In the absence of population-to-coherence coupling terms (the secular limit, $p=0)$, this matrix  further decouples into one $3\times 3$ and two $1\times 1$ subblocks.

Finally, the $4\times 4$ subblock of the Liouvillian matrix that acts on the bottom 4 elements of $\vec{\rho}$ ($\rho_{13}$, $\rho_{31}$, $\rho_{23}$, and  $\rho_{32})$ is given by
\begin{widetext}
\begin{equation}\label{LD4}
\bm{\mathcal{L}}_D^{(4)} = 
 \begin{pmatrix}
 -\frac{1}{2}(2r_1+r_2+\gamma_1)  & 0 &  \xi_{12}(p) & 0  \\
 0 & -\frac{1}{2}(2r_1+r_2+\gamma_1)  & 0 &  \xi_{12}(p)  \\
 \xi_{12}(p)  & 0 &  -\frac{1}{2}(r_1+2r_2+\gamma_1) & 0  \\
0  &   \xi_{12}(p) & 0 & -\frac{1}{2}(r_1+2r_2+\gamma_1)   \\
 \end{pmatrix}.
 \end{equation}
  \end{widetext}

Having obtained the matrix representation of the Liouvillian, we can now use Eq.~\eqref{aij_DM} to directly calculate the elements of the  Kossakowski matrix  for the V-system. To this end, we  first define $9$-dimensional complex vectors $\mathbf{f}_i$ by vectorizing  the SU$(N)$ basis matrices (see Sec. II) and then calculate the $\hat{\mathcal{L}_D}[\mathbf{F}_m]$ as matrix-vector products, i.e.,  $\hat{\mathcal{L}_D}[\mathbf{F}_m]=\bm{\mathcal{L}}_D \mathbf{f}_m$. These  matrix-vector products are subsequently ``devectorized'' and used in  matrix trace computations to produce the Kossakowski matrix elements, as prescribed by Eq.~\eqref{aij_DM}.


\subsection{$\Lambda$ system}

The density matrix of a three-level  $\Lambda$-system driven by incoherent radiation [see Fig.~1(b)] evolves in time according to the PSBR equation in the energy basis   \cite{Tscherbul:15b,Koyu:22,Ou:08}
\begin{align}\notag
\dot{\rho}_{13} &= -\frac{1}{2}(\gamma_1+\gamma_2+2r_1+r_2)\rho_{13} - \frac{p}{2}\sqrt{r_1r_2}\rho_{23},\\ \notag
\dot{\rho}_{23} &= -\frac{1}{2}(\gamma_1+\gamma_2+r_1+2r_2)\rho_{23} - \frac{p}{2}\sqrt{r_1r_2}\rho_{13},
\end{align}
\begin{align}\label{QME_lam}\notag
\dot{\rho}_{i i} &=-(r_i + \gamma_i)\rho_{ii} + (r_i+\gamma_i) \rho_{33}- p \sqrt{r_1r_2}\rho_{12}^R,  \\ 
\dot{\rho}_{1 2}&= -i\rho_{12}\Delta -\frac{1}{2}(r_1+r_2)\rho_{12}  \\ \notag
                &+{p}(\sqrt{r_1 r_2} +\sqrt{\gamma_1\gamma_2})\rho_{33}
                -\frac{p}{2}\sqrt{r_1r_2}(\rho_{11}+\rho_{2 2}), \\ \notag
                \end{align}                  
where $i=1,2$ denote the ground states of the $\Lambda$-system [see Fig.~1(a)], $3$ stands for the excited state, and the parameters $r_i$ and $\gamma_i$ have the same meaning as described below Eq.~\eqref{QME}.
 As in the V-system case considered in the previous section,  (i) the population-to-coherence coupling terms proportional to $p$ in Eq.~\eqref{QME} are responsible for the generation of noise-induced Fano coherences \cite{Koyu:22,Ou:08}, and (ii) the  dynamics of state populations and two-photon coherences is decoupled from that of one-photon coherences $\rho_{13}$ and $\rho_{23}$ in the  partial secular approximation.

Using the same definition of $\vec{\rho}$ is in Eq.~\eqref{DMvec} [note the different energy level notation in Figs.~1(a) and (b)] and noting that $\dot{\rho}_{33}=-\dot{\rho}_{11}-\dot{\rho}_{22}$, the $9\times 9$ Liouvillian supermatrix of the $\Lambda$ system  block-diagonalizes into  $5\times5$ and $4\times 4$ subblocks, $\bm{\mathcal{L}}_D = \bm{\mathcal{L}}_D^{(5)}\oplus \bm{\mathcal{L}}_D^{(4)}$, with 

 \begin{widetext}
\begin{equation}\label{LD5lam}
\bm{\mathcal{L}}_D^{(5)} = 
 \begin{pmatrix}
 - r_1 & 0 & r_1 + \gamma_1 & -\frac{p}{2}\sqrt{r_1 r_2} & -\frac{p}{2}\sqrt{r_1 r_2} \\
 0 & - r_2 & r_2+\gamma_2 & -\frac{p}{2}\sqrt{r_1 r_2} & -\frac{p}{2}\sqrt{r_1 r_2}  \\
r_1 &  r_2 & -(r_1+r_2+\gamma_1+\gamma_2) & p\sqrt{r_1 r_2}  & p\sqrt{r_1 r_2}   \\
-\frac{p}{2}\sqrt{r_1 r_2}   & -\frac{p}{2}\sqrt{r_1 r_2}  & -2\xi_{12}(p) & -\frac{1}{2}(r_1+r_2) & 0   \\
-\frac{p}{2}\sqrt{r_1 r_2}   & -\frac{p}{2}\sqrt{r_1 r_2}  & -2\xi_{12}(p) & 0 & -\frac{1}{2}(r_1+r_2)   \\
 \end{pmatrix},
 \end{equation}
  \end{widetext}
where $\xi_{12}(p)=-\frac{p}{2}(\sqrt{r_1 r_2}+\sqrt{\gamma_1\gamma_2})$. In the secular limit ($p=0$) this matrix further decouples into one $3\times 3$ and two $1\times 1$ subblocks, analogously to the V-system. Finally, the remaining $4\times 4$ subblock of the Liouvillian $\bm{\mathcal{L}}_D^{(4)} $ becomes  
 \begin{widetext}
\begin{equation}\label{LD4}
 \begin{pmatrix}
 -\frac{1}{2}(\gamma_1+\gamma_2 + 2r_1 + r_2)  & 0 &  -\frac{p}{2}\sqrt{r_1 r_2}  & 0  \\
 0 & -\frac{1}{2}(\gamma_1+\gamma_2 + 2r_1 + r_2)  & 0 &  -\frac{p}{2}\sqrt{r_1 r_2} \\
-\frac{p}{2}\sqrt{r_1 r_2}   & 0 &  -\frac{1}{2}(\gamma_1+\gamma_2 + r_1 + 2r_2) & 0  \\
0  &  -\frac{p}{2}\sqrt{r_1 r_2}   & 0 & -\frac{1}{2}(\gamma_1+\gamma_2 + r_1 + 2r_2)    \\
 \end{pmatrix}.
 \end{equation}
   \end{widetext}

\begin{figure*}[t]
\captionsetup[subfigure]{margin={0.6cm,0.0cm},font=normalsize}
    \centering
\subfloat[] {
  \includegraphics[width=0.65\columnwidth, trim = 0 0 0 0]{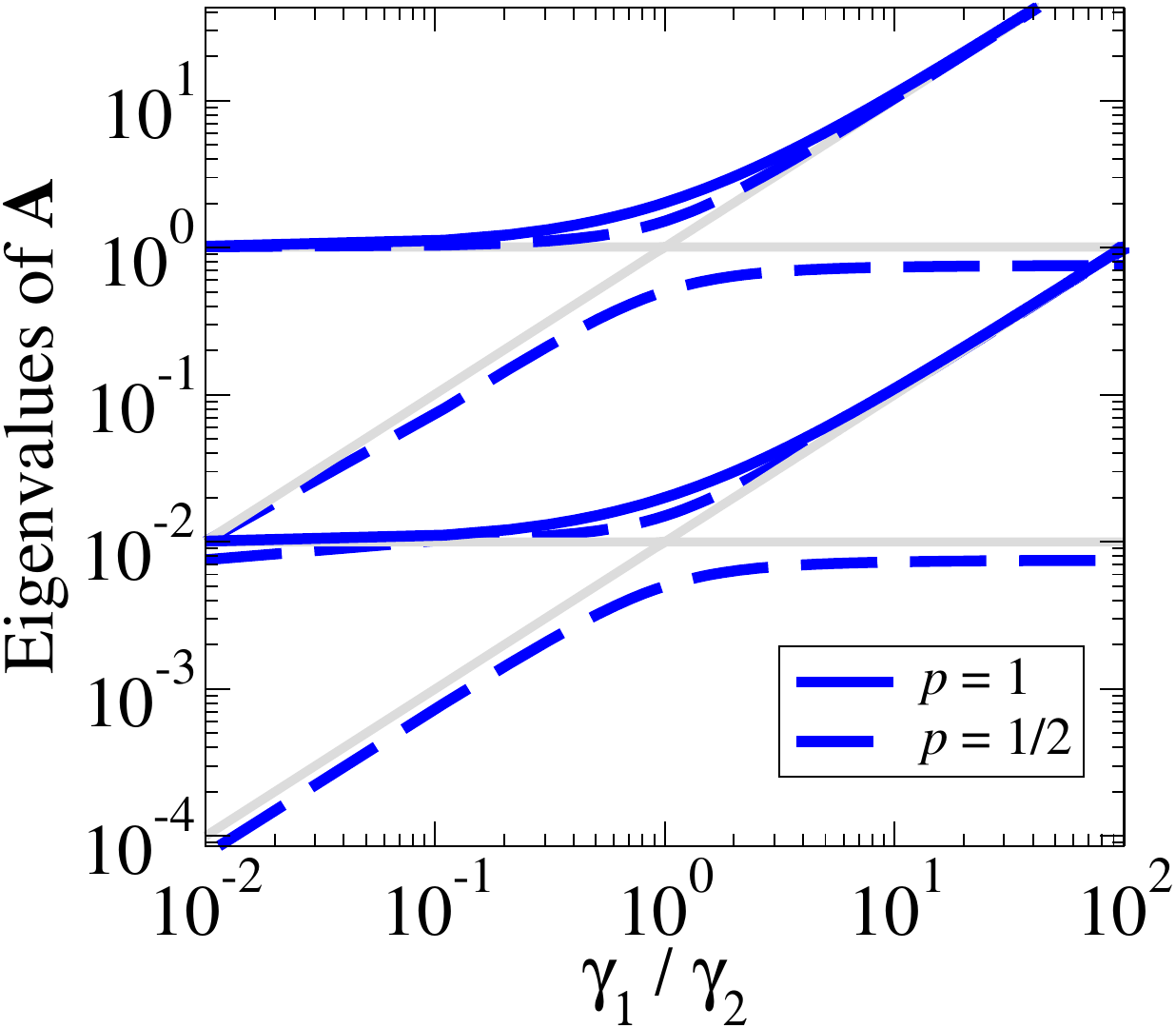}
}
\subfloat[]{
  \includegraphics[width=0.65\columnwidth,trim =  0 0 0 0]{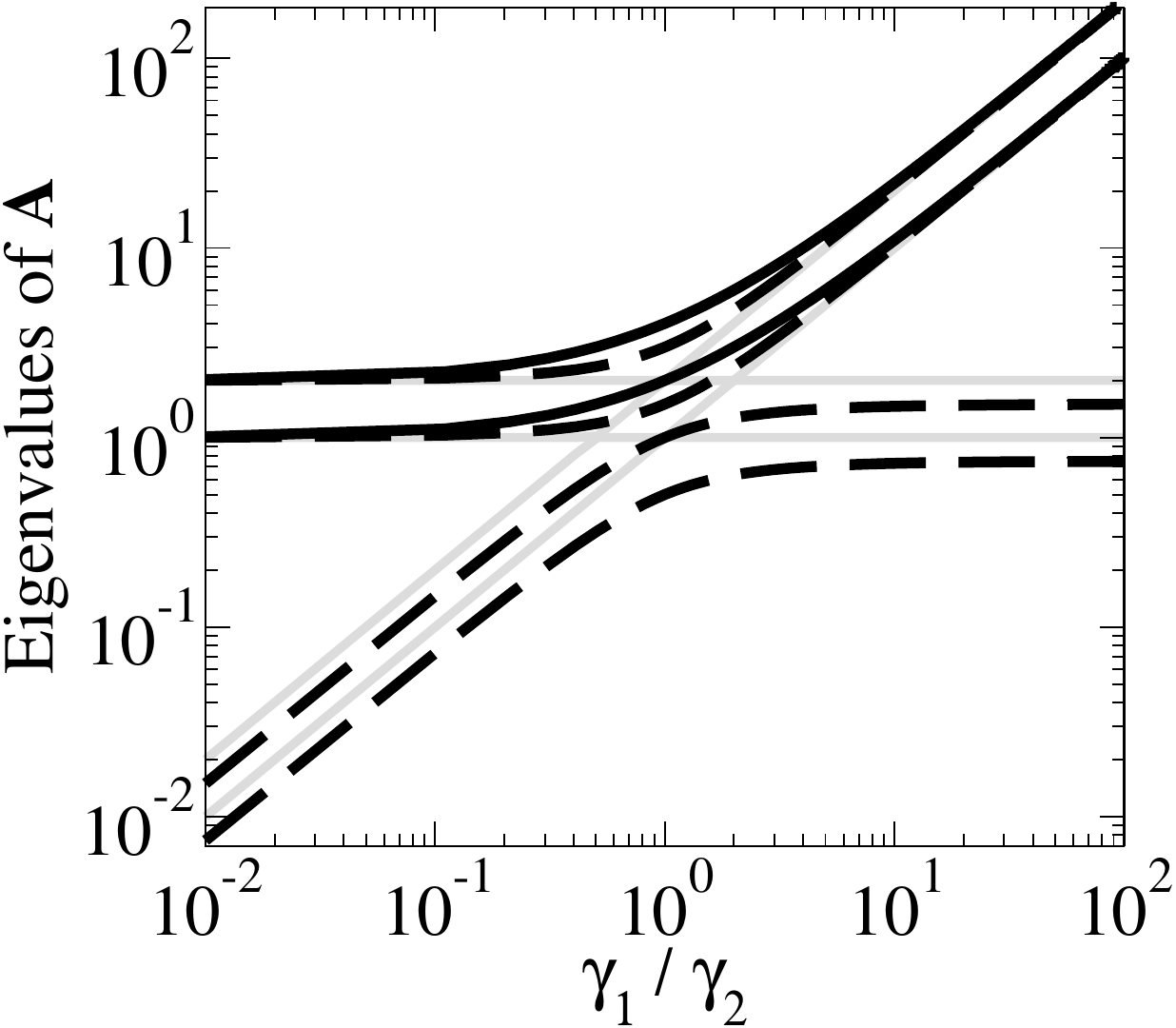}
}	
\subfloat[]{
  \includegraphics[width=0.65\columnwidth,trim =  0 0 0 0]{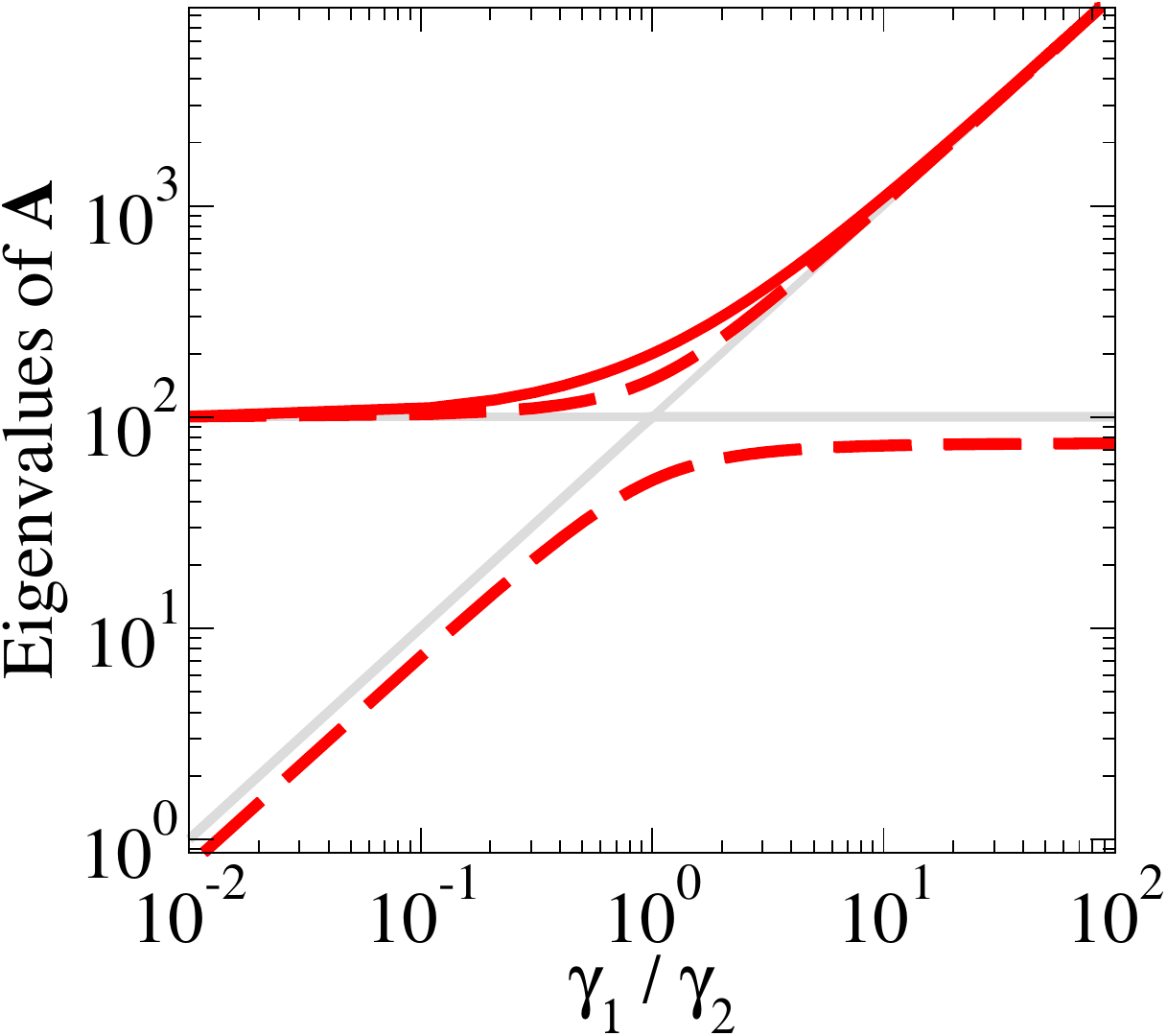}
}
\caption{Nonzero eigenvalues of the Kossakowski matrix of the PSBR equation for three-level V- and $\Lambda$ systems driven by isotropic incoherent radiation. The eigenvalues are plotted as a function of the ratio of spontaneous decay rates $\gamma_1/\gamma_2$ for collinear transition dipoles ($p=1$, solid lines), perpendicular transition dipoles ($p=0$, solid gray lines), and $p=1/2$ (dashed lines) in the weak [$\bar{n}=0.01$, panel (a)], intermediate [$\bar{n}=1$, panel (b)], and strong [$\bar{n}=100$, panel (c)] pumping regimes.}
	\label{fig:Vsys}
\end{figure*}

\subsection{Kossakowski matrices for three-level $\Lambda$ and V-systems driven by incoherent radiation}


Here, we examine the  eigenvalues of the Kossakowski matrix of the PSBR equations, which describes the quantum dissipative evolution of the $\Lambda$ and V-systems  driven by isotropic incoherent radiation. Using the approach outlined in Sec.~IIA, we found that the eigenvalues calculated for the V system are exactly the same as those for the $\Lambda$  system in all dynamical regimes spanned by the parameters $\gamma_1/\gamma_2$, $\bar{n}$, and $p$. This result  suggests that quantum dissipative dynamics of $\Lambda$ and V-systems driven by incoherent radiation can be described by essentially the same diagonal Lindblad equation, and thus can be expected to share closely similar features. This is an example of new insight afforded by the  Kossakowski matrix analysis, which is difficult to obtain by examining the (seemingly different) Liouvillians of the V- and $\Lambda$ systems.

 Figure~2 shows the eigenvalues of the Kossakowski matrix for the V and $\Lambda$-systems as a function of the ratio of spontaneous decay rates $\gamma_2/\gamma_1$, which quantifies the extent of asymmetry of two incoherent pumping transitions \cite{Dodin:16}.
  Importantly, we observe that the eigenvalues are non-negative across all the regimes studied, which include  the weak ($\bar{n}\ll 1$), intermediate  ($\bar{n}=1$), and strong  ($\bar{n}\gg 1$) pumping regimes, and transition dipole alignment parameters ranging from $p=1$ to $p=0$. This establishes the CP property of the quantum dynamical maps associated  with the PSBR equation for three-level V and $\Lambda$-systems driven by isotropic incoherent radiation, a central result of this work.

In the weak pumping limit relevant for photosynthetic sunlight-harvesting \cite{Tscherbul:14}, the incoherent pumping rate is much smaller than that of spontaneous decay $(\bar{n}\ll 1)$. Figure 2(a) shows that in this limit, there are only two nonzero eigenvalues for $p=1$, one of which  is much larger than the other. Both of the eigenvalues  increase monotonically with $\gamma_1/\gamma_2$. Four nonzero eigenvalues appear for $p<1$, also increasing steadily  with $\gamma_1/\gamma_2$. Interestingly,  the eigenvalues   at $\gamma_1=\gamma_2$ experience two genuine crossings in the secular limit $(p=0)$, which become avoided   for $p=1/2$.

 As shown in Figs. 2(b) and (c), the gap between the eigenvalues closes with increasing the pumping rate. At $\bar{n}=1$ the eigenvalues are within a factor of two of each other, and they become indistinguishable on the scale of Fig.~2(c) at $\bar{n}=100$. The qualitative trends of non-negativity and monotonic increase with the asymmetry parameter $\gamma_1/\gamma_2$ remain the same as in the weak pumping limit.

\section{Summary}

In summary, we have explored and implemented two direct methods to obtain the Kossakowski matrix of an arbitrary $N$-level Markovian QME from its Liouvillian matrix, thereby enabling straightforward CP tests of the associated quantum dynamical maps. We have applied these methods to calculate the Kossakowski matrices, and establish the CP property of the quantum dynamical evolution under the partial secular Bloch-Redfield QME of the three-level V and $\Lambda$ systems driven by isotropic incoherent radiation.

These approaches are general in the sense that they do not require one to specify an operator form of the QME, and could be applied in a straightforward manner to  many interesting QMEs, such as those, which occur in quantum thermodynamics  \cite{QuantumThermodynamicsBook,Alicki:79,Kosloff:19,Potts:24,Liu:21,Myers:22,Soret:22,Onishchenko:24,Scully:11}, photosynthetic  energy transfer \cite{Rebentrost:09,Cao:09,Gelin:11,Leon-Montiel:14,Tscherbul:18}, and spectroscopy \cite{YuenZhou:11,Mancal:12,YuenZhouBook,Fetherolf:17, Sayer:24}.  All that is required in the Liouvillian matrix, which can be  expressed in terms of relaxation and decoherence rates available either from experiments or via microscopic system-bath models  \cite{AlickiBook,Breuer:06}.  This will enable one to readily identify the regimes, in which the CP condition breaks down, and to apply the recently developed regularization methods   \cite{Abbruzzo:23,Abbruzzo:24prx,Pradilla:24} to restore CP evolution. An extension of this approach to non-Markovian QMEs \cite{Breuer:16,Vega:17,Chruscinski:22} could also be fruitful.

\section*{Acknowledgements}

The author is grateful to Michael  Hall for bringing Ref.~\cite{Hall:14} and Eq.~\eqref{aij_DM} to his attention, and to Suyesh Koyu, Amro Dodin,  Paul Brumer, Mikhail Lemeshko, and Johannes Feist for stimulating discussions.
This work was partially supported by the NSF CAREER program (grant No. PHY-2045681).

\appendix
\section{Derivation of Eq.~\eqref{avector_text_main} of the main text}

To derive Eq.~\eqref{avector_text_main} of the main text, we begin by using the identity  
$\mathbf{F}_m\mathbf{F}_n=\frac{1}{N}\mathbf{F}_0\delta_{mn} + \frac{i}{2}\sum_{l=1}^M z_{mnl}^*\mathbf{F}_l$, where $z_{mnl}=f_{mnl}+id_{mnl}$, and 
\begin{align}\label{fd}\notag
f_{ijk} &= -i \text{Tr}([\mathbf{F}_i,\mathbf{F}_j] \mathbf{F}_k), \\
d_{ijk} &=  \text{Tr}(\{\mathbf{F}_i,\mathbf{F}_j\} \mathbf{F}_k)
\end{align}
are the symmetric and antisymmetric structure constants of SU$(N)$ \cite{Lendi:87,AlickiBook}.  Expressing the elements of $\mathbf{R}$ and $\mathbf{k}$ in terms of those of $\mathbf{A}$ \cite{AlickiBook,Lendi:87,Pottinger:85}
\begin{align}\label{lineq1}
r_{sm}&=\sum_{i,k=1}^M \mathcal{T}_{sm,ik} a_{ik} \quad (s,m=1,\ldots,M) ,\\ \label{lineq2}
k_{s}&=\frac{i}{N} \sum_{i,k=1}^M a_{ik} f_{kls} \quad (s=1,\ldots,M),
\end{align}
where
\begin{equation}\label{ttensor}
\mathcal{T}_{sm,ik}=-\frac{1}{4}\sum_{l=1}^M [ (f_{mnl}+id_{mnl})f_{kls} + (f_{klm}-id_{klm}) f_{ils}) ],\\
\end{equation}
is an $M^2\times M^2$ transformation tensor  $\mathcal{T}$, which depends only on the symmetric and antisymmetric structure constants of SU$(N)$. The elements of  $\mathcal{T}$ are subject to a set of constraints,  which follow from the sum rules satisfied by the structure constants of SU$(N)$ as a compact semi-simple Lie group \cite{Macfarlane:68}. For example, one has, for $p=1,\ldots,M$
\begin{equation}\label{ttensor_rule1}
\sum_{i,k=1}^M \mathcal{T}_{sm,ik}f_{ikp}=i N d_{msp}/2.
\end{equation}

To express the elements of the Kossakowski matrix $a_{ik}$ in terms of the relaxation and decoherence rates ($r_{sm}$ and $k_s$),  we need to solve the system of linear equations \eqref{lineq1}-\eqref{lineq2}  for the  $a_{ik}$. Because of the constraints such as Eq.~\eqref{ttensor_rule1}, the complex square tensor $\mathcal{T}$ is not of full rank, and hence cannot be inverted, making the solution of linear equations \eqref{lineq1}-\eqref{lineq2} a nontrivial task. 

\begin{figure}
\captionsetup[subfigure]{margin={0.0cm,0.0cm},font=normalsize}
    \centering
\subfloat[] {
  \includegraphics[width=0.7\columnwidth, trim = 0 20 0 0]{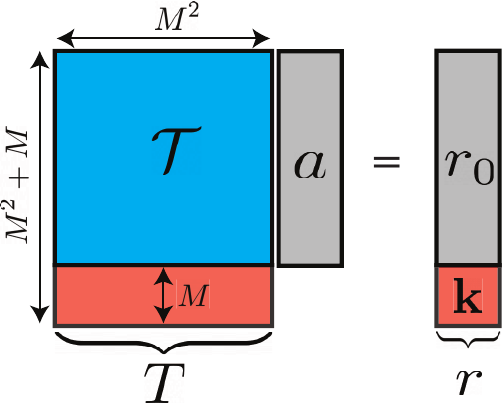}
}\\
\subfloat[]{
  \includegraphics[width=0.7\columnwidth,trim =  0 0 0 0]{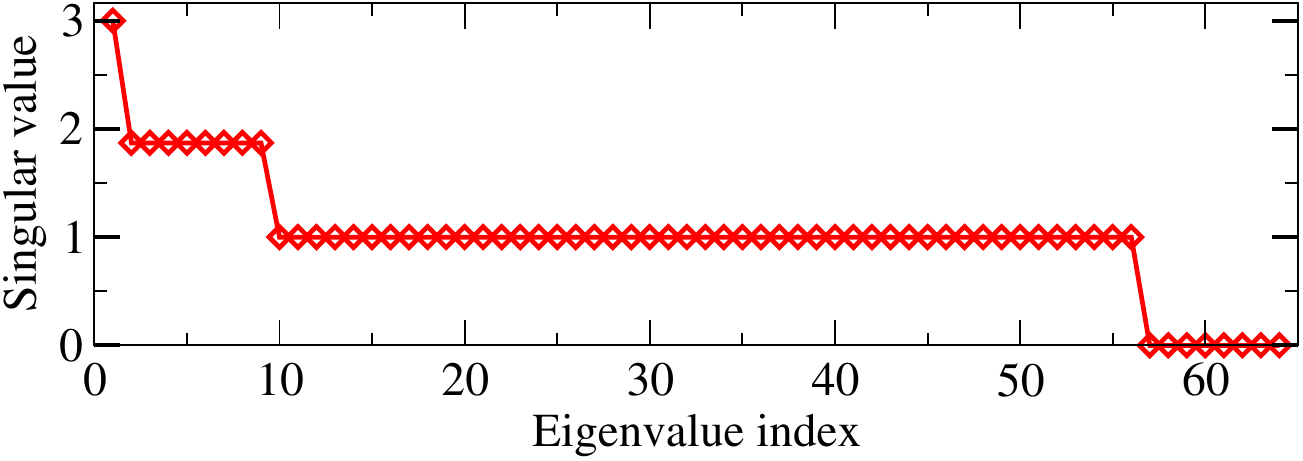}
}	
\caption{(a) Schematic structure of the matrix equation \eqref{Taug_lineq}. The complex rectangular matrix of SU$(N)$ structure constants $\bm{T}$ is composed of the complex $M^2\times M^2$ transformation tensor $\bm{\mathcal{T}}$ given by Eq.~\eqref{ttensor} augmented by the $M\times M^2$ matrix [see Eq.~\eqref{Taug}]. The right-hand side consists of the vectorized rate matrix $\bm{r}_0=\mathrm{vec}(\mathbf{R})$ augmented  at the bottom by the driving vector $\mathbf{k}$. (b) Singular values of the tensor $\bm{\mathcal{T}}$ for $N=3$. Note the presence of 8 zero singular values in the bottom right corner. The solid line is a guide to the eye.}
	\label{fig:mx_diagram}
\end{figure}

Note that Eq.~\eqref{lineq1} can be written as a superoperator  $\hat{\mathcal{T}}$ (or a process matrix \cite{Nielsen:10}), which acts on the Kossakowski matrix  to produce the rate matrix, $\hat{\mathcal{T}}\mathbf{A}=\mathbf{R}$. It is convenient to define vectorized (Liouville) representations \cite{Gyamfi:20} of the Kossakowski and rate matrices $\bm{a}=\text{vec}(\mathbf{A})$ and $\bm{r}_0=\text{vec}(\mathbf{R})$, where the components ${a}_\alpha$ and ${r}_{0\alpha}$ are uniquely related to the elements of $\mathbf{A}$ and $\mathbf{R}$, respectively, and $\alpha=\{ik\}=1,\ldots,M^2$ is a composite index.
We can now rewrite Eq.~\eqref{lineq1} in vectorized notation as 
$\bm{\mathcal{T}}\bm{a}=\bm{r}_0$, 
where the superoperator $\hat{\mathcal{T}}$  is represented by a square $M^2\times M^2$ tensor $\bm{\mathcal{T}}$ with elements $\mathcal{T}_{\alpha\beta}$ (note that the elements of ordinary $M\times M$ matrices, such as  $\mathbf{A}=\{a_{ik}\}$, are indexed by Latin letters).

Because the tensor $\bm{\mathcal{T}}$ is not of full rank (see above), the linear system $\bm{\mathcal{T}}\bm{a}=\bm{r}_0$ has infinitely many  solutions.
To obtain a unique solution, we need to specify $M$ additional constraints, which come in the form of Eq.~\eqref{lineq2}. To incorporate these constraints, we define a  rectangular tensor $\bm{T}$ with elements (the essential difference between the indices $s$ and $\beta=\{ik\}$ is noted above)
 \begin{align}\label{Taug}\notag
\qquad \quad T_{\alpha\beta} &= \mathcal{T}_{\alpha\beta} \, \quad (\alpha,\beta=1,\ldots, M^2), \\
\qquad \quad T_{M^2+s,\beta} &= \frac{i}{3} f_{sik} \, \, (s=1,\ldots,M).
\end{align}
As illustrated in Fig.~3(a), this definition amounts to augmenting the square tensor $\bm{\mathcal{T}}$ with additional rows populated by the structure constants $(i/3)f_{sik}$ [see Eq.~\eqref{Taug}].
The system of linear equations \eqref{lineq1}-\eqref{lineq2} can then be recast in tensor form  
 \begin{equation}\label{Taug_lineq}
\bm{{T}}\bm{a}=\bm{r},
\end{equation}
where $\bm{r}$ is an augmented vectorized rate matrix with  ${r}_\alpha = r_{0\alpha}=r_{ik}$ for $\alpha=1,\ldots,M^2$ and ${r}_{M^2+s} = k_{s}$ [see Eqs.~\eqref{lineq1}-\eqref{lineq2}].

The  $(M^2+M) \times M^2$ complex rectangular tensor $\bm{T}$  has rank $M^2$, and thus  the  vectorized Kossakowski matrix $\bm{a} =\text{vec}(\mathbf{A})$ can be obtained as 
\begin{equation}\label{avector}
\bm{a}=\bm{{T}}^{+} \bm{r}, 
\end{equation}
where 
\begin{equation}\label{Tsupermatrix}
\bm{{T}}^{+} = (\bm{T}^\dag \bm{T})^{-1} \bm{T}^\dag
\end{equation}
is the Moore-Penrose pseudo-inverse of $\bm{T}$ \cite{Moore:20,Penrose:55,Barata:12} (the superscript $^+$ should not be confused with that for Hermitian conjugation, $\dag$).
Equation \eqref{avector} gives the Kossakowski matrix $\bm{a}=\text{vec}(\mathbf{A})$ in terms of the relaxation and decoherence rates of a general-form QME encapsulated in $\bm{r}$. 

To obtain the rate matrix $\mathbf{R}$ and the driving vector $\mathbf{k}$ for the three-level V-system, we first define the eight-component coherence vector $\mathbf{v}= \sqrt{2}( \rho_{12}^R, -\rho_{12}^I,\frac{1}{2}(\rho_{11}-\rho_{22}),\rho_{13}^R,-\rho_{13}^I,\rho_{23}^R,-\rho_{23}^I,\frac{1}{\sqrt{12}}(\rho_{11}+\rho_{22}-2\rho_{33}) )^T$
  in terms of the density matrix elements using the SU(3) decomposition \eqref{SUNdecomp}, and then recast the QME \eqref{QME} in the form of Eq.~\eqref{coherence_vec}, as described in Appendix B. The resulting $8\times 8$ $\mathbf{R}$ matrix, whose elements are expressed in terms of $r_i$, $\gamma_i$, and $p$ (see Appendix B), is vectorized to form the 64-component vector $\bm{r}_0$. The latter is  subsequently augmented with the 8-component $\mathbf{k}$ vector to obtain the 72-component vector $\bm{r}$ on the right-hand side of Eq.~\eqref{Taug_lineq} (see above and Fig.~1). 
  
  The elements of the transformation tensor $\bm{T}$ are calculated using Eq.~\eqref{Taug} using the structure constants for the $\{F_i\}$ basis of SU(3) \cite{AlickiBook}. 
 Next, the Moore-Penrose pseudo-inverse of $\bm{T}$ is calculated using Eq.~\eqref{Tsupermatrix}.  To verify the correctness of this procedure, we calculated the singular value decomposition of the $64\times 64$ complex tensor $\bm{\mathcal{T}}$ shown by the blue square  in Fig.~3(a)
 \begin{equation}\label{SVD}
\bm{\mathcal{T}}= \mathbf{U} \mathbf{\Sigma} \mathbf{V}^\dag,
\end{equation}
 where $\mathbf{\Sigma}$ contains the singular values of $\bm{\mathcal{T}}$ and $\mathbf{U}$ and $\mathbf{V}$ are unitary matrices.
  Figure 1(b) shows that $8$ of the 64 singular values of $\bm{\mathcal{T}}$ are zero. This is consistent with the expectation that the rank of $\bm{\mathcal{T}}$ is $M^2-M$, with $M$ additional constraints needed to define $\bm{T}$ (note that $M=N^2-1=8$ for the three-level system).
We also verified that Eq.~\eqref{Taug_lineq} is consistent with Eq.~(164) of Ref.~\cite[p.~69]{AlickiBook}. To our knowledge, the authors of Refs.~\cite{AlickiBook,PottingerThesis}  expressed the Liouvillian matrix elements in terms of the $a_{ik}$ for $N=3$, but did not perform the inversion step  accomplished in this work, which gives  the $a_{ik}$ in terms of the Liouvillian matrix elements, i.e., relaxation and decoherence rates.

\section{PSBR equation for the coherence vector of the V-system}

To derive the equations of motion for the coherence vector, we begin with
 the PSBR equations \eqref{QME} for the  density matrix of the V-system  ($N=3$)  driven by incoherent radiation  \cite{Kozlov:06,Tscherbul:14,Tscherbul:15b,Dodin:16}.
We first separate these  equations into their real and imaginary parts, and then form their linear combinations to match the components of the coherence vector \eqref{SUNdecomp}. For example, the equation of motion for ${\rho}_{13}$ becomes
 \begin{align}\label{sm_QMEinterm}\notag
\dot{\rho}_{13}^R &= -\frac{1}{2}(2r_1+r_2+\gamma_1)\rho_{13}^R - \frac{p}{2}(\sqrt{r_1r_2}+\sqrt{\gamma_1\gamma_2})\rho_{23}^R,\\ 
\dot{\rho}_{13}^I &= -\frac{1}{2}(2r_1+r_2+\gamma_1)\rho_{13}^I - \frac{p}{2}(\sqrt{r_1r_2}+\sqrt{\gamma_1\gamma_2})\rho_{23}^I,
\end{align}
Multiplying these equations by $\pm\sqrt{2}$ and using Eq.~\eqref{SUNdecomp}, we get 
 \begin{multline}\label{sm_QME45}
\dot{v}_4 = \sqrt{2}\dot{\rho}_{13}^R = -\frac{1}{2}(2r_1+r_2+\gamma_1) \sqrt{2}\rho_{13}^R  \\
-\frac{p}{2}(\sqrt{r_1r_2}+\sqrt{\gamma_1\gamma_2}) \sqrt{2}\rho_{23}^R \\
= -\frac{1}{2}(2r_1+r_2+\gamma_1) v_4 - \frac{p}{2}(\sqrt{r_1r_2}+\sqrt{\gamma_1\gamma_2})  v_6, \\ 
\end{multline}
and
 \begin{multline}
\dot{v}_5 = -\sqrt{2} \dot{\rho}_{13}^I 
= -\frac{1}{2}(2r_1+r_2+\gamma_1) (-\sqrt{2})\rho_{13}^I  \\
 -\frac{p}{2}(\sqrt{r_1r_2}+\sqrt{\gamma_1\gamma_2})\rho_{23}^I \\
= -\frac{1}{2}(2r_1+r_2+\gamma_1) v_5  -\frac{p}{2}(\sqrt{r_1r_2}+\sqrt{\gamma_1\gamma_2}) v_7,
\end{multline}

Proceeding in the same way for the other components of $\mathbf{v}$, we obtain the dissipative part of the QME as
\begin{equation}\label{sm_coherence_vec}
\dot{\mathbf{v}}(t) = \mathbf{R}\mathbf{v}(t) + \mathbf{k}, 
\vspace{0.1cm}
\end{equation}
where the nonzero elements of the $8\times 8$ rate matrix $\mathbf{R}$ are given by
 \begin{align}\label{Rnz} \notag
R_{11} &= -\bar{r} - \bar{\gamma}, \,\, R_{18} = -\frac{p\sqrt{3}}{3}(3r_{12}+\gamma_{12}), \\ \notag
R_{22} &=  -\bar{r} - \bar{\gamma}, \\ \notag
R_{33} &= -\bar{r} - \bar{\gamma}, \,\, R_{38} = -\frac{p\sqrt{3}}{3}(3r_{12}+\gamma_{12}), \\ \notag
R_{44} &= -\frac{1}{2}(2r_1+r_2+\gamma_1), \,\, R_{46} = -\frac{p}{2}(r_{12}+\gamma_{12}), \\ \notag
R_{55} &=  -\frac{1}{2}(2r_1+r_2+\gamma_1), \,\, R_{57} =-\frac{p}{2}(r_{12}+\gamma_{12}) , \\ \notag
R_{64} &=   -\frac{p}{2}(r_{12}+\gamma_{12}), \,\, R_{66} =-\frac{1}{2}(r_1+2r_2+\gamma_2)  , \\ \notag
R_{75} &=  -\frac{p}{2}(r_{12}+\gamma_{12}), \,\, R_{77} =-\frac{1}{2}(r_1+2r_2+\gamma_2)   , \\ \notag
R_{81} &=  -p\sqrt{3}(r_{12}+\gamma_{12}), \,\, R_{88}=-3\bar{r} - \bar{\gamma}, \\ 
R_{83} &=-\frac{\sqrt{3}}{2}[-(r_1+\gamma_1)+r_2+\gamma_2)],
\end{align}
 where $\gamma_{12}=\sqrt{\gamma_1\gamma_2}$, $r_{12}=\sqrt{r_1 r_2}$, $\bar{r}=\frac{1}{2}(r_1+r_2)$, and $\bar{\gamma}=\frac{1}{2}(\gamma_1+\gamma_2)$. Finally, the driving vector $\mathbf{k}$ is given by 
  \begin{equation}\label{sm_kvector}
\mathbf{k}= 
 \begin{pmatrix}
-\frac{p\sqrt{2}}{3}\gamma_{12} \\
0 \\
 -\frac{1}{3\sqrt{2}}(\gamma_1-\gamma_2)\\
0 \\
0 \\
0 \\
0 \\
-\frac{1}{\sqrt{6}}(\gamma_1+\gamma_2)
 \end{pmatrix}.
 \end{equation}

\end{document}